\documentstyle[12pt]{article}
\begin{document}

\title{The equivalence principle in classical mechanics and quantum 
mechanics\footnote{gr-qc/9810087, December 22, 1998}}

\author{\normalsize{Philip D. Mannheim} \\
\normalsize{Department of Physics,
University of Connecticut, Storrs, CT 06269} \\ 
\normalsize{mannheim@uconnvm.uconn.edu} \\}

\date{}

\maketitle

\begin{abstract}
We discuss our understanding of the equivalence principle in both classical 
mechanics and quantum mechanics. We show that not only does the equivalence
principle hold for the trajectories of quantum particles in a background 
gravitational field, but also that it is only because of this that the
equivalence principle is even to be expected to hold for classical particles 
at all. 
\end{abstract}

\subsection*{1~~~The Equivalence Principle in Classical Mechanics}

While the equivalence principle stands at the very heart of general relativity,
there appear to be some aspects of it that are not quite as secure as they 
might be. In particular, while there is no disputing the fact that classical 
geodesics in a background gravitational field exhibit the equivalence 
principle, the question of whether the motions of real physical systems can 
explicitly be associated with such geodesics is actually a logically 
independent issue. Moreover, so also is the further question of what is 
supposed to happen when the physical systems are to be described by quantum 
mechanics, a situation which has actually been explored experimentally in the 
landmark Colella-Overhauser-Werner (COW) study 
\cite{Overhauser1974,Colella1975,Greenberger1979,Werner1994} of a 
quantum-mechanical beam of neutrons traversing an interferometer located in an 
external gravitational field. In this paper we shall examine both of these 
issues to show that not only is the equivalence principle actually found to 
hold for quantum-mechanical particles, but that classical-mechanical 
particles actually inherit the classical-mechanical equivalence principle from 
them.  

While the standard road to the $\hbar=0$ classical-mechanical equivalence 
principle is of course completely familiar (see e.g. \cite{Weinberg1972}), it 
is nonetheless pedagogically instructive to quickly recall the steps. Suppose 
we begin with a standard, free, spinless, relativistic,  classical-mechanical 
Newtonian particle of non-zero kinematic mass $m$ moving in flat spacetime 
according to the special relativistic generalization of Newton's second law of 
motion                                 
\begin{equation}
m{d^2\xi^{\alpha} \over d \tau ^2} =0~~,~~~R_{\mu \nu \sigma \tau}=0 
\label{1}
\end{equation}
where $d\tau=(-\eta_{\alpha \beta}d\xi^{\alpha} d\xi^{\beta})^{1/2}$ is
the proper time and $\eta_{\alpha \beta}$ is the flat spacetime metric, and 
where we have indicated explicitly that the Riemann tensor is (for the moment) 
zero. Now let us transform to an arbitrary coordinate system $x^{\mu}$. Using  
the definitions 
\begin{equation}
\Gamma ^{\lambda}_{\mu \nu} = {\partial x^{\lambda} \over \partial
\xi^{\alpha}} {\partial ^2\xi^{\alpha} \over \partial x^{\mu} \partial
x^{\nu}}~~~~,~~~~g_{\mu \nu}= {\partial \xi^{\alpha} \over \partial
x^{\mu}} {\partial \xi^{\beta} \over \partial x^{\nu}}
\eta_{\alpha \beta} \label{2}
\end{equation}
we find directly (see e.g. Ref. \cite{Weinberg1972,Mannheim1993}) that the 
invariant proper time takes the 
form $d\tau=(-g_{\mu \nu}dx^{\mu} dx^{\nu})^{1/2}$, while the 
equation of motion of Eq. (\ref{1}) gets rewritten as
\begin{equation}
m \left( {D^2x^{\lambda} \over D\tau^2}\right) \equiv
m \left( {d^2x^{\lambda} \over d\tau^2}
+\Gamma^{\lambda}_{\mu \nu} 
{dx^{\mu} \over d\tau}{dx^{\nu } \over d\tau} \right) 
= 0~~~,~~~R_{\mu \nu \sigma \tau}=0 
\label{3}
\end{equation}
which serves to define $D^2x^{\lambda}/D\tau^2$. As derived, Eq. (\ref{3}) so 
far only holds in a strictly flat spacetime with zero Riemann curvature tensor, 
and indeed Eq. (\ref{3}) is only a covariant rewriting of the special 
relativistic Newtonian second law of motion, i.e. it covariantly describes 
what an observer with a non-uniform velocity in flat spacetime sees, with
the $\Gamma ^{\lambda}_{\mu \nu}$ term emerging as an inertial, coordinate 
dependent force. While the four-velocity $dx^{\lambda}/d\tau$ is a general 
contravariant vector, its ordinary derivative $d^2x^{\lambda}/d\tau^2$ (which 
samples adjacent points and not merely the point where the four-velocity itself 
is calculated) is not, and it is only $D^2x^{\lambda}/D\tau^2$ which transforms 
as a general contravariant four-acceleration, and it is thus only this 
particular four-vector on whose meaning all (accelerating and 
non-accelerating) observers can agree. 

As regards the generalization of Eq. (\ref{3}) to include a coupling of the 
non-zero mass particle to gravity, the great insight of Einstein was to then 
realize that in a non-flat spacetime if the gravitational field emerged  
purely from the Christoffel symbol $\Gamma^{\lambda}_{\mu \nu}$, Eq. (\ref{3}) 
would then be replaced by the non-flat
\begin{equation}
m \left( {d^2x^{\lambda} \over d\tau^2}
+\Gamma^{\lambda}_{\mu \nu}
{dx^{\mu} \over d\tau}{dx^{\nu } \over d\tau} \right)=0~~~,~~~R_{\mu 
\nu \sigma \tau}\neq0. 
\label{4}
\end{equation}                 
Equation (\ref{4}) achieves two things - it establishes the metric as the 
gravitational field in the first place, and, further, it specifies how a 
classical, relativistic, non-zero mass test particle is to couple to gravity. 
Moreover, given this Eq. (\ref{4}), no less than three forms of the 
classical-mechanical equivalence principle\footnote{Collectively these three 
complementary forms constitute (see e.g. \cite{Will1993}) what is known as the 
weak equivalence principle, the principle which is usually invoked in order to 
fix the laws of motion for particles undergoing free fall in an external 
gravitational field.} then become apparent: (i) since both of the two terms on 
the left-hand side of Eq. (\ref{4}) have the same coefficient $m$, the equality 
of the inertial and (passive) gravitational masses is automatically secured; 
(ii) since the mass parameter $m$ only appears as an irrelevant overall 
multiplier in Eq. (\ref{4}), the trajectories associated with the integration 
of Eq. (\ref{4}) are then independent of the masses of test particles; (iii) 
precisely because the Christoffel symbol is not a general coordinate tensor, it 
is always possible to find some general coordinate system in which all the 
components of $\Gamma^{\lambda}_{\mu \nu}$ can be made to vanish at some 
particular 
point,\footnote{Under the transformation $x^{\prime \lambda} = x^{\lambda}+
\frac{1}{2}x^{\mu}x^{\nu}(\Gamma^{\lambda}_{\mu \nu})_{P}$, the primed 
coordinate Christoffel symbols $(\Gamma^{\prime \lambda}_{\mu \nu})_{P}$ will 
be forced to vanish at the point $P$, independent in fact of how curved the 
Riemann tensor at that same point $P$ might be, i.e. regardless of how strong 
the gravitational field at $P$ might actually be, and even regardless of what 
particular covariant gravitational equation of motion is actually used to fix 
the Christoffel symbols in the first place.} with it then being possible 
(cf. Eqs. (\ref{1}) and (\ref{3})) to simulate the gravitational field at such 
a point by an accelerating coordinate system in flat spacetime. Since
this same geodesic equation of motion can directly be obtained as the  
stationarity condition $\delta I_T/\delta x_{\lambda}=0$ on the 
relativistically covariant action
\begin{equation}
I_T=-mc\int d\tau                                                                               
\label{5}
\end{equation}               
under variation with respect to the coordinates of the particle itself, we 
thus see that a particle whose action is in fact the point test particle 
action $I_T$ will then automatically move geodesically in a background 
classical gravitational field and will then necessarily obey (all three of the 
above forms of) the classical-mechanical equivalence principle.

In order to distinguish the above three forms of the equivalence principle 
(something necessary for the discussion of the quantum-mechanical case which we 
give below), we note that the very writing of $I_T$ in the form 
$-mc\int d\tau$ entails that $I_T$ only involves one mass parameter $m$. Thus 
there are not two independent mass scales available to even permit independent 
parameterizations of the inertial and gravitational masses in the first place. 
With this one mass parameter also only acting as an overall multiplier 
in $I_T$, both the first and second forms of the classical-mechanical 
equivalence principle are thus seen to be explicit consequences of using the 
action $I_T$. However, the third form of the equivalence principle given above, 
namely our ability to remove the Christoffel symbols at any chosen point, is 
actually a property of the geometry itself independent of the existence or 
otherwise of any such $I_T$, and thus needs to be considered in and of itself. 

To explicitly illustrate this specific point it is convenient to consider the 
geometry near the surface of a static, spherically symmetric source of (active) 
gravitational mass $M$ and radius $R$ such as the earth or a star. For such 
sources the geometrical line element can be written as 
$d\tau^2=B(r)c^2dt^2-dr^2/B(r)-r^2d\Omega$ where $B(r)=1-2MG/c^2r$. If we erect
a Cartesian coordinate system $x=r$sin$\theta$cos$\phi$, 
$y=r$sin$\theta$sin$\phi$, $z=r$cos$\theta - R$ at the surface of the earth, 
then, with $z$ being normal to the earth's surface, to lowest order in 
$x/R,~y/R,~z/R,~MG/c^2R~(=gR/c^2)$ the Schwarzschild line element is then 
found \cite{Moreau1994} to take the form
\begin{equation}
d\tau^2=[1-a(z)]c^2dt^2-dx^2-dy^2-[1+a(z)]dz^2-b(xdx+ydy)dz
\label{6}
\end{equation}
where $a(z)=2g(R-z)/c^2$ and $b=4g/c^2$. For the metric of Eq. (\ref{6}) we 
can then calculate the Christoffel symbols near the surface of the earth with 
Eq. (\ref{4}) then yielding
\begin{equation}
\ddot{x}=0~,~ \ddot{y}=0~,~ \ddot{t}+(2g/c^2)\dot{t}\dot{z}=0~,~
\ddot{z}+(g/c^2)(2\dot{x}^2+2\dot{y}^2-\dot{z}^2)+g\dot{t}^2=0~,
\label{7}
\end{equation}
where the dot denotes differentiation with respect to the proper time 
$\tau$. For trajectories for which the initial velocity $v$ is in the 
horizontal $x$ direction, to lowest order in $x/R,~y/R,~z/R,~gR/c^2$ 
Eq. (\ref{7}) then reduces to 
\begin{equation}
\ddot{x}=0~,~ \ddot{y}=0~,~ \ddot{t}=0~,~
\ddot{z}+2gv^2/c^2+g=0~,
\label{8}
\end{equation}
(on using $ \ddot{t}=0$ to set $ \dot{t}=1$), 
with the motion thus being equivalent to that of an acceleration in flat 
spacetime. Consequently, we anticipate that it must be possible to 
transform the line element of Eq. (\ref{6}) to the flat coordinate 
$d\tau^2=c^2dt^{\prime 2}-dx^{\prime 2}-dy^{\prime 2}-dz^{\prime 2}$, 
something which we indeed readily achieve via the transformation
\begin{eqnarray}
x^{\prime}=x~,~y^{\prime}=y~,~t^{\prime}=t(1-gR/c^2+gz/c^2)~,
\nonumber \\
z^{\prime}=z(1+gR/c^2-gz/2c^2)+gt^2/2+g(x^2+y^2)/c^2.
\label{9}
\end{eqnarray}
While we thus show that near its surface the earth's gravity indeed acts the 
same way as an acceleration in flat spacetime,\footnote{The $x=0$, $y=0$, $z=0$ 
origin of the unprimed system obeys $z^{\prime}=gt^{\prime 2}/2$ in the primed 
system, with the Cartesian primed system straight line 
$x^{\prime}=vt^{\prime}$, $y^{\prime}=0$, $z^{\prime}=0$ taking the unprimed 
form $x=vt(1-gR/c^2)$, $y=0$, $z=-g(1+2v^2/c^2)t^2/2$ in accord with 
Eq. (\ref{8}).} 
some caution is needed in trying to interpret this result.\footnote{In passing
we note that while this same analysis could of course also be made for other
metrics as well, it turns out to give an instructive surprise when applied
to the specific metric $d\tau^2=B(r)c^2dt^2-dr^2/B(r)-r^2d\Omega$ where  
now $B(r)=1-2MG/c^2r+\gamma r$, viz. the specific metric obtained in the 
alternate conformal gravity theory described in \cite{Mannheim1998a} and 
references therein. Specifically, in this case in the weak gravity limit near 
the surface of the earth (i.e. $\gamma R$ also small), the metric is found to 
reduce to the same generic form as given in Eq. (\ref{6}) where now  
$a(z)=2g(R-z)/c^2 -\gamma (R+z)$ and $b=4g/c^2-2\gamma$. The resulting metric 
is also flat and it can be directly brought to the Cartesian form by 
the transformation $x^{\prime}=x$, $y^{\prime}=y$, 
$z^{\prime}=z(1+gR/c^2-gz/2c^2-\gamma R/2 -\gamma z/4)
+(g+\gamma c^2/2)t^2/2+(g/c^2-\gamma /2)(x^2+y^2)$, 
$t^{\prime}=t(1-gR/c^2+gz/c^2+\gamma R/2+\gamma z/2)$, a transformation under 
which the unprimed system origin obeys 
$z^{\prime}=(g+\gamma c^2/2)t^{\prime 2}/2$ and under which the primed system 
straight line $x^{\prime}=vt^{\prime}$, $y^{\prime}=0$, $z^{\prime}=0$ takes 
the form $x=vt(1-gR/c^2+\gamma R/2)$, $y=0$, 
$z=-g(1+2v^2/c^2)t^2/2-\gamma c^2(1-2v^2/c^2)t^2/4$ in 
the unprimed system. In the unprimed 
coordinate system the metric yields trajectories of the form 
$\ddot{x}=0$, $\ddot{y}=0$, $\ddot{t}=0$,
$\ddot{z}+2(g-\gamma c^2/2)v^2/c^2+g+\gamma c^2/2=0$. Thus in the 
non-relativistic $v=0$ case the motion is described by 
$\ddot{z}+g+\gamma c^2/2=0$, while in the relativistic $v=c$ case we instead
obtain $\ddot{z}+3g-\gamma c^2/2=0$. Thus, as already noted in 
Refs. \cite{Walker1994,Edery1998}, the effect of the $\gamma$ term is 
opposite in these two limits. Thus the fact that a potential may be attractive 
for non-relativistic motions does not in and of itself mean that it must 
therefore also be attractive for light, with the $v^2/c^2$ type terms not only 
modifying the magnitude of the effect of gravity (something already found by 
Einstein following his various attempts to calculate the gravitational bending 
of light), but even being able to modify the sign of the effect as well. Thus 
in general we see that even after fixing the sign of the numerical coefficient 
of a gravitational potential term once and for all, such a potential is then 
not necessarily always attractive.}  Specifically, since the metric of Eq. 
(\ref{6}) can 
be transformed to a flat metric, despite its explicit dependence on $g$ the 
metric of Eq. (\ref{6}) must therefore be flat, with explicit calculation 
directly confirming the vanishing of the Riemann tensor associated with Eq. 
(\ref{6}) in this low order. Thus we need to ascertain what happened to the 
non-trivial curvature which the full Schwarzschild metric is known to possess, 
a curvature which is associated with an explicitly non-zero 
Riemann tensor even in lowest order in $g$. The answer to 
this puzzle is that while the Christoffel symbols are first order derivative 
functions of the metric, the Riemann tensor is a second order derivative 
function. Thus to get the lowest non-trivial term in the Riemann tensor we need 
to expand the metric to second order in $x/R,~y/R,~z/R$. Since a first order 
expansion suffices for the Christoffel symbols, we thus see that there is a 
mismatch between orders of expansion of the Christoffel symbols and the Riemann 
tensor. Hence a first order study of the geodesics is simply not sensitive to 
the curvature, and thus we see why the equivalence principle not only works for 
weak gravity near the surface of the earth, but why in fact it even has to do 
so.\footnote{In fact the Christoffel symbols are the only non-trivial first 
order derivative functions of the metric which are available since the 
covariant first order derivative of the metric just happens to vanish 
identically in a Riemannian geometry. Since the Christoffel symbols themselves 
are not covariant tensors, the identical vanishing of 
$g^{\mu \nu}_{\phantom{\mu\nu};\nu}$ entails that there is no non-trivial 
covariant first order derivative function of the metric at all. Consequently 
lowest order study of the inertial properties of geodesics cannot be sensitive 
to any non-inertial, coordinate independent, covariant properties of the metric 
such as its curvature.} 

In order to underscore this last point, we note that the essence of Eq. 
(\ref{4}) is that it asserts that the coupling of test particles to gravity is
purely inertial, with there being no direct coupling of the particle to any
non-inertial, coordinate independent quantities such as the Riemann curvature.
However, in principle, Eq. (\ref{3}) admits of covariant generalizations other 
than that given in Eq. (\ref{4}). For instance, equations of motion such as the 
fully covariant 
\begin{eqnarray}
m\left( {d^2x^{\lambda} \over d\tau^2}
+\Gamma^{\lambda}_{\mu \nu}
{dx^{\mu} \over d\tau}{dx^{\nu } \over d\tau} \right)
=-\kappa_1 R^{\beta}_{\phantom {\beta} \beta} 
\left( {d^2x^{\lambda} \over d\tau^2}
+\Gamma^{\lambda}_{\mu \nu}
{dx^{\mu} \over d\tau}{dx^{\nu } \over d\tau} \right)
\nonumber \\
-\left(\kappa_1 R^{\beta}_{\phantom {\beta} \beta ;\alpha}
+\kappa_2 R_{\alpha \beta}                                                        
{dx^{\beta } \over d\tau}+ 
\kappa_3 R_{\alpha \beta \gamma \delta}{dx^{\beta } \over d\tau}
{dx^{\gamma } \over d\tau}S^{\delta}\right)                                                       
\left( g^{\lambda \alpha}+                                                      
{dx^{\lambda} \over d\tau}                                                      
{dx^{\alpha} \over d\tau}\right)
\label{10}
\end{eqnarray}                 
($\kappa_1$, $\kappa_2$ and $\kappa_3$ are appropriate constants and 
$S^{\delta}$ is an appropriate spin vector) also reduce back to the flat 
spacetime Eq. (\ref{3}) in the absence of curvature; with their departure
from Eq. (\ref{4}) in the presence of curvature being due to terms which 
enable the particle to exchange energy and momentum with the gravitational 
field as it propagates in that field, terms which can thus be regarded as a 
gravitational analog of the electromagnetic Lorentz force.\footnote{The 
$\kappa_1$ dependent term, for instance, can be derived as the variation with 
respect to $x_\lambda$ of the action $I=-\kappa_1\int d\tau R^{\beta}_{\phantom 
{\beta} \beta}$, and can thus be considered as a curvature dependent analog of 
the electromagnetic action $I=e\int d\tau A_{\mu}(dx^{\mu}/d\tau)$.}  
Moreover, in the presence of these additional terms, the motion of a test 
particle would then depend on its mass, with it being impossible to ever 
remove these additional terms at any point by any clever choice of coordinates. 
Despite the fact that Eq. (\ref{10}) would lead to a non-null result for an 
Eotvos type experiment performed in a strong enough gravitational field such 
as, for instance, the Ricci non-flat one found near the surface of a charged 
or a radiating black hole, we note that since the $\kappa_i$ 
dependent terms in Eq. (\ref{10}) are all proportional to the Riemann tensor, 
they are all second order in $x/R,~y/R,~z/R$. Consequently, weak gravity tests 
of the equivalence principle for non-zero mass particles in the small 
$x/R,~y/R,~z/R,~MG/c^2R$ limit are somewhat insensitive to their possible 
presence (at least for some range of values of the $\kappa_i$), with such weak 
gravity tests of the equivalence principle thus not requiring the explicit use 
of Eq. (\ref{4}) where such curvature dependent terms are explicitly assumed to 
be absent.\footnote{For actual practical purposes we note that since all three 
of the $\kappa_i$ dependent terms just happen to vanish for spinless particles 
propagating in a Ricci flat geometry, weak gravity tests near the surface of 
the earth would anyway be completely insensitive to their possible presence.} 
This may be just as well, since no proof appears to have ever been given in the 
standard gravity literature which would actually enable us to explicitly 
exclude equations of motion such as that of Eq. (\ref{10}) for real, non-zero 
mass classical particles in the real world.

In order to emphasize the full import of this last remark, we note that for 
many people the equivalence principle is actually regarded as an axiom or as
a postulate which is considered to be separate from general relativity itself 
and which is thus to be imposed as an addition to general relativity. Indeed, 
with test particles being defined as particles which move on geodesics, the 
(weak) equivalence principle postulate may be regarded as the assumption (or 
requirement) that for gravitational purposes real particles can be treated as 
test particles. Now since the equivalence principle predates full general 
relativity, and since the equivalence principle was so instrumental in its 
eventual development, historically there was thus a time when the equivalence 
principle did have an independent existence. However, now that general 
relativity is so well established, we need to ask whether the equivalence 
principle actually is or is not a consequence of general relativity; and in a 
sense, one would only consider the equivalence principle to be logically 
independent of general relativity if one was prepared to contemplate that 
results such as the (weak equivalence principle establishing) null Eotvos 
result could be valid even while the full (strong equivalence principle) 
covariant apparatus of general relativity might somehow not be. Moreover, 
there even appears to be a shortcoming in how the weak equivalence principle 
is ordinarily formulated in the first place. Specifically, it is generally 
viewed as a "covariantization" prescription, viz. take the equations of motion 
of special relativity and simply replace everything that appears in those 
equations (such as the metric, ordinary derivatives and ordinary accelerations) 
by their general covariant analogs (such as the contravariant four-acceleration 
$D^2x^{\lambda}/D\tau^2$ of Eqs. (\ref{3}) and (\ref{4})). Thus, since point 
particles, and equally light rays, move on straight lines in gravity free flat 
spacetime, covariantization then requires particles and rays to be just as 
geodesic when they propagate in a background gravitational field. However, as 
explicitly exhibited in Eq. (\ref{10}), such a prescription is not the most 
general one that is permitted by the requirement that the equations of motion 
which are to describe the motions of particles in a background gravitational 
field be general coordinate invariant. Thus the central question for real (as 
opposed to test) particles is to what degree their dynamics is constrained by
the requirement that they are to couple covariantly to gravity, and whether 
such a constraint in and of itself does or does not lead to their having to 
obey the weak equivalence principle (viz. Eq. (\ref{4})) without further 
assumption. It is this apparently not yet explored issue then which we shall 
specifically study in this paper,\footnote{While a great deal of attention has
of course been given to the weak equivalence principle in the literature, 
nonetheless the primary thrust has generally been to ask whether there might be 
any possible non-covariant departure from the geodesic motion exhibited in 
Eq. (\ref{4}) (the issue of preferred frames being a typical example), rather 
than whether there could or should be any possible covariant departure such as
that exhibited in Eq. (\ref{10}).} and as we shall see, the question actually 
relates to the dynamics of particles rather than to any constraints that 
geometry might impose on their kinematics.     

As we have so far seen, our above discussion actually raises no less than 
three separate issues: (i) to what extent are real physical 
systems actually describable by the test particle action $I_T$ - and if not 
to what extent are they actually then geodesic (i.e. to what extent do we need 
to guard against the possible presence of explicit curvature dependent terms
such as those exhibited in Eq. (\ref{10})); (ii) how do we demonstrate that 
light waves propagate geodesically (for massless particles $mc \int d\tau$ 
vanishes); and (iii) to what extent does the discussion carry over to 
quantum-mechanical systems, systems which exhibit both particle and wave 
aspects. Interestingly, through the use of the quantum-mechanical study we make 
below, we will find that we will then be able to establish an explicit 
equivalence principle for massive particles in the classical limit.

In order to specifically address the issue of the extension of the equivalence 
principle to the massless light wave case, we note that, unlike the particle 
case, this time we have to begin with a classical wave equation. While we shall 
see below that our analysis will actually need modifying in the presence of 
waves with polarization, for simplicity we shall initially restrict our 
discussion to massless scalar classical waves, viz. waves which obey the 
covariant minimally coupled curved space scalar field wave equation (viz. the
covariantized form of the flat space massless Klein-Gordon equation)
\begin{equation}
S^{\mu}_{\phantom{\mu};\mu}=0~~,~~~R_{\mu \nu \sigma \tau} \neq 0
\label{10a}
\end{equation}
where $S^{\mu}$ denotes the contravariant 
derivative $\partial S/\partial x_{\mu}$ of the scalar field $S(x)$. If we 
set $S(x)=$ exp$(iT(x))$, the eikonal 
phase $T(x)$ is then found to obey the equation 
$T^{\mu}T_{\mu}-iT^{\mu}_{\phantom{;\mu};\mu}=0$, a condition which reduces to 
$T^{\mu}T_{\mu}=0$ in the short wavelength limit. From the associated 
condition $T^{\mu}T_{\mu ;\nu}=0$ it then follows that 
$T^{\mu}T_{\nu ;\mu}=0$. Since normals to the wavefronts obey the eikonal   
relation $T^{\mu}=dx^{\mu}/dq=k^{\mu}$ where $q$ is a convenient affine 
parameter which measures distance along the normals and where $k^{\mu}$ is the 
wave vector of the wave, we thus obtain (see e.g. \cite{Mannheim1993}) 
$k^{\mu}k^{\nu}_{\phantom{\nu} ;\mu}=0$, a condition which we recognize as 
being exactly the massless particle geodesic equation, with rays then 
precisely being found to be geodesic in the eikonal limit. Thus we see that 
once given the standard minimally coupled Klein-Gordon equation, then not only 
do we indeed obtain geodesic motion, but, additionally, we explicitly find 
that no additional curvature dependent terms such as the ones exhibited in 
Eq. (\ref{10}) are in fact then generated. Thus unlike the non-zero mass case, 
we see that for the massless wave case the very imposition of the minimally 
coupled Klein-Gordon equation eliminates any possible ambiguity in how massless 
waves might couple to gravity, and forces them to be strictly geodesic (in weak
or strong gravity and both near to or far from a gravitational source for that 
matter). Since the discussion given earlier of the coordinate dependence of the 
Christoffel symbols was purely geometric, we thus see that once rays such as 
light rays are geodesic, they immediately obey the third form of the 
equivalence principle given earlier, 
with phenomena such as the gravitational bending of light then immediately 
following.\footnote{According to Eqs. (\ref{7}) and (\ref{8}), an observer in 
Einstein's elevator would not be able to tell if a light ray (for light the dot 
symbol in Eq. (\ref{7}) denotes differentiation with respect to the affine 
parameter $q$) is falling downwards under gravity or whether the elevator is 
accelerating upwards.} Thus we see that the third form of the equivalence 
principle has primacy over the first and second ones for light (indeed light 
has no inertial mass or gravitational mass to begin with), and that geodesic 
motion need not be intimately tied to the classical test particle action
$I_T$ at all. As we thus can anticipate, since quantum-mechanical systems obey 
wave equations, the discussion of the quantum-mechanical equivalence principle 
should be expected to be closer in spirit to the discussion associated with 
light rather than with that associated with test particles, a point which we 
explore below.     

However, before leaving the classical-mechanical equivalence principle, it is 
instructive to note that not only is there an Eq. (\ref{10}) type ambiguity in 
the response of non-zero mass particles to an external gravitational field, 
whenever these same particles act as gravitational sources there again is an 
analogous such ambiguity, one which then affects the specific gravitational 
fields these sources are capable of producing. Specifically, we note that it 
is conventional in applications of general relativity to classical macroscopic 
non-zero mass sources to simply take the energy-momentum tensor of typical 
gravitational sources to be kinematic perfect fluids, viz. of the form 
$T_{kin}^{\mu \nu}=(\rho+p)U^{\mu}U^{\nu} +p g^{\mu \nu}$ where $\rho$ and 
$p$ are the fluid energy density and pressure. The motivation for doing this is 
that (i) this form provides a covariant generalization of the flat spacetime 
perfect fluid form which is known to work extremely well for 
non-gravitational interactions, and (ii) the covariant conservation of this
same $T_{kin}^{\mu \nu}$ precisely leads to free fall for the particles in the 
fluid in the pressure free case, i.e. precisely to geodesic 
motion.\footnote{Thus if the particles in a composite object such as a 
planet are all in free fall in the gravitational field of the sun, then the 
center of mass of the planet will be in free fall too.} While this emergence 
of geodesic motion is of course highly desirable, it has led to the assertion 
that macroscopic sources then are in fact perfect fluids. However, this 
quite widespread assertion is unwarranted, since while the above discussion 
does show that the condition $(T_{kin}^{\mu \nu})_{;\nu}=0$ is indeed 
sufficient to give geodesic motion, it does not follow that $T_{kin}^{\mu \nu}$ 
is then necessarily the complete gravitational source $T^{\mu \nu}$. Indeed, 
$T_{kin}^{\mu \nu}$ need not be the full covariant source $T^{\mu \nu}$ of 
gravity, since the covariant conservation of $T^{\mu \nu}=T_{kin}^{\mu \nu}
+T_{extra}^{\mu \nu}$ will also give geodesic motion in the presence of any 
$T_{extra}^{\mu \nu}$ which is itself separately covariantly 
conserved.\footnote{Typical possible candidates for such a 
$T_{extra}^{\mu \nu}$ would be a tensor which transforms as $g^{\mu \nu}$ or 
one which transforms as the Einstein tensor $R^{\mu \nu}-
\frac{1}{2}g^{\mu \nu}R^{\alpha}_{\phantom{\alpha} \alpha}$, a tensor whose 
possible presence or absence in the full gravitational $T^{\mu \nu}$ is simply 
not ascertainable via studies of non-gravitational interactions.} Moreover, 
since $T_{kin}^{\mu \nu}$ only involves the excitations of the one-particle 
sector out of the vacuum, it is the only piece of the full gravitational 
$T^{\mu \nu}$ which is measurable in non-gravitational physics, a regime which 
(in contrast to gravity) is only sensitive to changes in energy and not to 
their zero.\footnote{In passing we note that the notorious cosmological 
constant problem derives from the difficulty inherent in locating the position 
of none other than this very zero. Thus, absent a resolution of the 
cosmological constant problem, it is simply impossible to assess whether  
any $g^{\mu \nu}$ type term may or may not be present in the full 
gravitational $T^{\mu \nu}$.} Thus just as we noted in our discussion of Eq. 
(\ref{10}), in the presence of gravity it is perfectly reasonable to anticipate 
the emergence of explicit curvature dependent terms in $T^{\mu \nu}$ in 
gravitational sources in which gravitational binding plays a direct dynamical 
role.\footnote{In passing we note that it turns out \cite{Mannheim1993} that if 
the full $T^{\mu \nu}$ source of gravity is locally conformal invariant, then 
even while the spontaneous breakdown of the conformal symmetry then explicitly 
induces a specific $T_{extra}^{\mu \nu}$, no exchange of energy and momentum 
between it and $T_{kin}^{\mu \nu}$ is actually found to occur, with 
$T_{kin}^{\mu \nu}$ still being covariantly conserved in this particular case.} 
And even while such explicit curvature dependent effects may be negligible in 
weak gravity sources such as normal stars, our ability to neglect them there in 
no way entails their absence in the strong gravity limit associated with
collapsed stars or black holes. Thus, without a demonstration that such
explicit curvature dependent terms are absent (or negligible), it is not yet
warranted to assert that macroscopic sources are describable by perfect
fluids at all. Hence not only is the response of matter to an external 
gravitational field not yet fully understood in classical physics, but 
neither is the mechanism by which classical matter sets up such gravitational 
fields in the first place. 

\subsection*{2~~~The Equivalence Principle in Quantum Mechanics}

While our above classical study of the response of non-zero mass particles to 
a gravitational field relied heavily on the use of the action 
$I_T=-mc\int d\tau$, and while this action even leads (in the presence of a 
gravitational source of mass $M$) to the non-relativistic Newtonian Lagrangian 
$L=T-V$ where $T=m_iv^2/2$, where $V=-m_gMG/r$, (and where $m_i=m_g=m$), it is 
important to note that, even though its variation would have led to the 
selfsame non-relativistic Newtonian Law of Gravity, nonetheless, the Newtonian 
Hamiltonian $H=T+V$ was not actually explicitly encountered in our above 
discussion, with it actually being somewhat peripheral to it.\footnote{In 
passing we note that in general relativity even though the energy-momentum 
tensor is locally covariantly conserved, nonetheless its global integrals
(such as $Q^{\mu}=\int d^3x (-g)^{1/2}T^{0\mu}$) are not necessarily constants 
of the motion in a general curved spacetime (and not even contravariant 
vectors in general), thus making it difficult to even define a 
curved spacetime Hamiltonian in the general case, or to know whether $Q^{0}$ 
might be able to serve as one in some specific one. Now, for the restricted 
case of the test particle action $I_T=-mc\int d\tau$, the associated $Q^{\mu}$ 
is readily calculated through use of the energy-momentum tensor $T^{\mu \nu}=   
2(-g)^{-1/2}\delta I_T/ \delta g_{\mu \nu}=mc (-g)^{-1/2}
\int d\tau \delta^4 (x-y(\tau)) (dy^{\mu}/d\tau) (dy^{\nu}/ d\tau)$, and 
is found to take the simple, suggestive form $Q^{\mu}=mc dx^{\mu}/d\tau$. 
However, explicit evaluation of the non-relativistic limit of this particular 
$Q^{\mu}$ in the curved geometry where $g_{00}=-(1-2MG/c^2r)$ is then found to 
yield $cQ^{0} \rightarrow mc^2+mv^2/2+mMG/r \sim mc^2+T-V$, an expression 
which is not of our desired $T+V$ form. On the other hand, in this same 
geometry the quantity $-cQ_{0}$ does in fact reduce to $mc^2+mv^2/2-mMG/r$ and 
thus is nicely of the $T+V$ form. In order to determine which one, if either, 
of these two particular quantities is to be identified as the energy of the 
particle, we recall that in classical mechanics there is actually a second, 
entirely different way to define the energy, viz. via the quantities 
$R_{\mu}=\partial I_M^{ST}/ \partial x^{\mu}$, where the derivatives act on the 
end point $x^{\mu}$ of the integral of some general matter action $I_M^{ST}$ as 
calculated in the specific stationary path which minimizes $I_M$. Since $I_M$ 
is always a general coordinate scalar, the $R_{\mu}$ always transform as a 
covariant four-vector no matter how complicated a function $I_M$ might be, so 
that $E=-cR_{0}$ can nicely serve as a well-defined energy even in curved 
spacetime. As we show in detail below, explicit evaluation of $R_{\mu}$ in the 
situation where $I_M$ is the test particle action $I_T$ then yields 
$R_{\mu}=\partial I_T^{ST}/ \partial x^{\mu}=mc dx_{\mu}/d\tau$, i.e. just the 
one which does in fact lead to the requisite $E=mc^2+T+V$. Now it is important 
to stress that in flat spacetime the two quantities $cQ^{0}$ and $-cQ_{0}$ 
actually do coincide, with it being only in curved spacetime that there is any 
difference between these covariant and contravariant quantities. Thus it is 
through the explicit use of the stationary action that we are able to identify 
the covariant $mc dx_{\mu}/d\tau$ rather than the contravariant 
$mc dx^{\mu}/d\tau$ as the appropriate energy-momentum vector (i.e. the energy 
is conjugate to the contravariant time $t=x^{0}/c$ and can thus be defined as 
the derivative of the stationary action with respect to $x^{0}$). Thus, as we 
see, our very ability to introduce an appropriate non-relativistic Hamiltonian 
at all requires us to first 
formulate an appropriate fully relativistic theory, an issue we explicitly 
take care of in the following.} Despite this, in the non-relativistic quantum 
case it is precisely the Hamiltonian which plays the most direct role, with the 
Schrodinger equation
\begin{equation}
i\hbar{\partial \psi \over \partial t}=-{\hbar^2 \over 2m_i}\nabla^2 \psi
+V(r)\psi
\label{11}
\end{equation}
following directly as the quantization of $H=T+V=E$. Thus we immediately need 
to ask whether it is in fact legitimate to use this familiar and very tempting 
quantization prescription in the case where $V$ is not just any arbitrary 
potential energy but is in fact precisely the gravitational $V(r)=-m_gMG/r$. 
Thus we now have to ask all over again whether it is in fact legitimate to set 
$m_i=m_g=m$ in the quantum-mechanical Eq. (\ref{11}), and note immediately that 
even if we are allowed to do so, nonetheless, the mass parameter $m$ no longer 
appears as the overall multiplier found in the classical Eq. (\ref{4}). 
Consequently, we see that no matter what the quantum-mechanical status of the 
first form of the equivalence principle given above, the second form 
immediately fails in quantum mechanics, with the solutions to the Schrodinger 
equation very much depending on the mass $m$. Consequently, at this point all 
three of our forms of the classical equivalence principle now become 
questionable in quantum mechanics, and so it is to this issue to that we now 
turn. 

In order to bypass having to address the issue of the quantization of 
the non-relativistic Newtonian $H$, one could instead start with the massive 
flat spacetime Klein-Gordon equation 
\begin{equation}
S^{\mu}_{\phantom{\mu};\mu}-(mc/\hbar)^2S=0~~,~~~R_{\mu \nu \sigma \tau}=0
\label{11a}
\end{equation}
and make the general coordinate transformation of Eq. 
(\ref{9}) from flat Cartesian coordinates to those associated with the 
(still flat) metric of Eq. (\ref{6}). In this way Greenberger and 
Overhauser \cite{Greenberger1979} found that a subsequent non-relativistic 
reduction of Eq. (\ref{11a}) in such coordinates then led precisely to the 
Schrodinger equation of Eq. (\ref{11}) near the surface of the earth. Under 
such a procedure the presence of only one mass parameter $m$ in the initial 
Klein-Gordon equation then entailed the equality $m_i=m_g=m$ in the 
resulting Schrodinger equation (with the gravitational potential energy 
near the surface of the earth then being given by $V=mgz$ ). However, since  
all of this analysis was made in flat spacetime, it served only to show that 
quantization of $H=T+V$ is equivalent to simulating gravity by an 
acceleration in flat spacetime, and thus did not address the question of 
whether in quantum mechanics it is actually legitimate to simulate true 
curvature by such an acceleration in the first place.

To address the specific issue of the actual quantum-mechanical status of the 
third form of the equivalence principle, we must instead look at a fully 
covariant analysis of the above non-zero mass Klein-Gordon equation 
\begin{equation}
S^{\mu}_{\phantom{\mu};\mu}-(mc/\hbar)^2S
=0~~,~~~R_{\mu \nu \sigma \tau} \neq 0
\label{12}
\end{equation}
as covariantized to curved spacetime. And 
indeed we find \cite{Mannheim1998b} that in the background gravitational field 
of the earth, viz. $d\tau^2=B(r)c^2dt^2-dr^2/B(r)-r^2d\Omega$ where 
$B(r)=1-2MG/c^2r$, setting  
$S(x)=$ exp$(-imc^2t/\hbar)\psi(x)$ then 
explicitly leads under non-relativistic reduction to 
\begin{equation}
i\hbar{\partial \psi \over \partial t}+{\hbar^2 \over 2m}\nabla^2 \psi
={mc^2 \over 2}[B(r)-1]\psi=-{mMG \over r}\psi,
\label{13}
\end{equation}
to thus not only lead us directly to Eq. (\ref{11}) and to not only directly
enforce $m_i=m_g=m$, but also to show that on restricting the
Schwarzschild metric to the weak gravity Eq. (\ref{6}) near the surface of 
the earth, we are then able to directly recover the inertial result presented 
in Ref. \cite{Greenberger1979}. Thus we show that our third (purely geometric) 
form of the equivalence principle does survive quantum mechanics, with a 
gravitational field still being simulatable by an acceleration in flat 
spacetime even for quantum-mechanical systems.

Since our above discussion recovers the equivalence principle without any 
reference to $I_T=-mc\int d\tau$, we see that, just as in our earlier 
discussion of the classical-mechanical Klein-Gordon equation, the 
test particle action $I_T$ is again found to be somewhat 
peripheral to the equivalence principle, with the heart of the issue being
not the first or the second forms of the equivalence principle at all, but
rather the primacy of the third form instead.\footnote{From the point of view 
of quantum mechanics, the equality of $m_i$ and $m_g$ should really be regarded 
as an equality of the inertial ($h/m_iv$) and gravitational ($h/m_gv$) de 
Broglie wavelengths, with the quantum-mechanical equivalence principle being
interpretable as the statement that neutron beams interfere in horizontal and
vertical interferometers with one and the same de Broglie wavelength $h/mv$; 
and with the general rule being that rather than coupling primarily to mass, 
gravity (itself a field theory) couples first and foremost to wavelength (and 
then only subsequently - via second quantization - to mass).}
Now while we have just seen that we did not need to introduce $I_T$, it is 
nonetheless possible to make some contact with it. Specifically, if we make 
the substitution $S(x)=$ exp$(iP(x)/\hbar)$ in the quantum-mechanical curved 
spacetime Klein-Gordon equation given as Eq. (\ref{12}), we then obtain 
\begin{equation}
P^{\mu}P_{\mu}+m^2c^2=i\hbar P^{\mu}_{\phantom{\mu}; \mu}.
\label{14}
\end{equation}
In the eikonal or in the small $\hbar$ approximation the 
$i\hbar P^{\mu}_{\phantom{\mu};\mu}$ 
term can be dropped, so that the phase $P(x)$ is then seen to obey the
purely classical condition 
\begin{equation}
g_{\mu \nu}P^{\mu}P^{\nu}+m^2c^2=0,
\label{15}
\end{equation}
a condition which we immediately recognize as the covariant Hamilton-Jacobi 
equation of classical mechanics, an equation whose solution is known to be the 
stationary classical action $\int p_{\mu}dx^{\mu}$ between relevant end 
points.\footnote{Viz. the action as calculated in that particular path which 
minimizes it.} In the eikonal approximation then we can thus 
identify the wave phase $P(x)$ as $\int p_{\mu}dx^{\mu}$, with the phase 
derivative $P^{\mu}$ then being given as the particle momentum 
$p^{\mu}=mcdx^{\mu}/d\tau$, a four-vector momentum which obeys   
\begin{equation}
p^{\mu}p_{\mu}+m^2c^2=0,
\label{16}
\end{equation}
viz. the familiar fully covariant particle energy-momentum 
relation.\footnote{In passing we note that in a static, weak gravitational 
background metric with $g_{00}=-(1-2MG/c^2r)$, use of the Hamilton-Jacobi 
equation thus enables us to identify the non-relativistic classical energy as 
$E=-c\partial P/\partial x^{0} \rightarrow mc^2+mv^2/2-mMG/r$. Thus while we 
see that we can indeed identify the non-relativistic Hamiltonian as $H=T+V$, 
we are able to do so only after having first obtained Eqs. (\ref{15}) and 
(\ref{16}), i.e. only after the quantum-mechanical problem associated with the 
fully covariant Klein-Gordon equation has already been solved. Moreover, it 
is important to emphasize, that even while we thus are able to make contact 
with the classical Hamiltonian, we were only able to do so in the eikonal 
approximation where the $i\hbar P^{\mu}_{\phantom{\mu};\mu}$ term is dropped.
However, since it was only the non-relativistic reduction of the full 
Klein-Gordon equation which led to the Schrodinger equation of Eq. (\ref{11}),
we see that the $i\hbar P^{\mu}_{\phantom{\mu};\mu}$ term does (in principle) 
contribute non-relativistically ($\nabla^2 \psi=\psi \bar {\nabla} (ln \psi) 
\cdot \bar {\nabla} (ln \psi)+\psi\nabla^2 (ln \psi) $) and thus can play 
a role in fixing the quantum-mechanical Hamiltonian, with the gravitational 
Schrodinger equation actually going beyond the eikonal approximation in the 
long wavelength limit.} Covariant differentiation of Eq. (\ref{16}) 
immediately leads to the classical massive particle geodesic equation 
$p^{\mu}p^{\nu}_{\phantom{\nu} ;\mu}=0$, to thus recover the well known 
(wave-particle duality) result that quantum-mechanical rays move on classical 
geodesics, i.e. that the center of a quantum-mechanical wave packet follows the 
classical trajectory.\footnote{In Refs. \cite{Mannheim1998b,Mannheim1996} this 
result was utilized to analyze the gravitationally induced quantum interference 
detected in the COW neutron beam interferometry experiment. The interested 
reader may find some further, complementary discussion of the implications 
of the COW experiment for the quantum-mechanical equivalence principle in 
Ref. \cite{Lammerzahl1996}.} Further, since we  
may also reexpress the stationary $\int p_{\mu}dx^{\mu}$ as $-mc\int d \tau$, 
we see that we can also identify the quantum-mechanical eikonal phase as 
$P(x)=-mc\int d\tau$, to thus nicely enable us to make contact with $I_T$ after 
all. Though we see that the classical action $I_T$ does thus play a role, it is 
important to realize that even though this classical action is indeed a 
solution to the classical Hamilton-Jacobi equation, we were only able to arrive 
at Eq. (\ref{15}) after first imposing the equation of motion of Eq. 
(\ref{12}), i.e. only after variation of the Klein-Gordon action had already 
been made, with only the stationary classical action (viz. $I_T^{ST}$) 
actually being a solution to the Hamilton-Jacobi equation. Thus in the 
quantum-mechanical case, just as noted for the Hamiltonian $H=T+V$, we see that 
the action $I_T=-mc\int d\tau$ is a part of the solution, i.e. the output, 
rather than being part of the input.\footnote{Thus we cannot appeal to $I_T$ 
to put particles on geodesics, since we already had to put them on geodesics 
in order to get to $I_T$ in the first place.} Thus unlike the situation in 
the classical-mechanical case, in the quantum-mechanical case we never need to 
assume the existence of any point particle action $I_T$ at all. Rather we need 
only assume the existence of equations such as the standard Klein-Gordon 
equation, with the eikonal approximation then precisely putting particles onto 
classical geodesics just as desired. 

Now while we have just seen that use of the standard Klein-Gordon equation does 
indeed lead to geodesic motion, we still need to address the fact that the 
standard minimally coupled Klein-Gordon equation (viz. the one obtained by 
writing the flat spacetime one covariantly) is not in fact the most general one 
that could be used in curved spacetime because of possible direct non-inertial 
couplings to 
curvature once the Riemann tensor is non-zero, couplings that could potentially 
lead us to the kind of (weak) equivalence principle violating terms exhibited 
in Eq. (\ref{10}). To explicitly address this issue, as well as the issue of 
how the discussion might need modifying in the presence of non-zero spin, we 
note first that with the fundamental fields of nature all being thought to be 
describable by renormalizable field theories, since such second quantized 
fields are to obey second quantized wave equations, the matrix elements of 
these second quantized fields between the vacuum and the one-particle state 
will then typically obey first quantized wave equations just like the 
first-quantized Klein-Gordon equation (and its spin non-zero analogs) which was 
studied above. Moreover, with all the fundamental field theories of nature even
being thought to be local gauge field theories, since such theories have no 
dimensionful coupling constants (and no intrinsic mass scales either if all 
masses are to generated dynamically) these theories then contain no explicit 
fundamental dimensionful parameters which might serve as the $\kappa_i$ 
parameters exhibited in Eq. (\ref{10}). Thus as far as the couplings of 
fundamental fields to curvature tensors such as the Riemann tensor are 
concerned, the only permissible such ones would have to involve dimensionless 
couplings, with the scalar field wave equation of Eq. (\ref{12}) for instance 
then being generalized to the non-minimally coupled
\begin{equation}
S^{\mu}_{\phantom{\mu};\mu}-(mc/\hbar)^2S 
+(\xi /6)R^{\alpha}_{\phantom {\alpha} \alpha}S
=0~~,~~~R_{\mu \nu \sigma \tau} \neq 0
\label{17}
\end{equation}
where $\xi$ is a dimensionless parameter, a parameter which, incidentally, 
takes the specific value $\xi=1$ should the theory possess an additional local 
conformal invariance above and beyond its local gauge 
invariance.\footnote{In fact once local conformal invariance is invoked, it 
alone is sufficient to unambiguously fix the couplings of all fields to 
gravity, with the $\xi=1$ term being the only direct Riemann curvature tensor 
dependent one possible.} 

Now even though (massless) fermions couple to gravity only through the fermion 
spin connection $\Gamma_{\mu}(x)$, so that the massless curved 
space Dirac equation $i\gamma^{\mu}(x)\nabla_{\mu}\psi(x)=0$ (where 
$\nabla_{\mu}=\partial_{\mu}+\Gamma_{\mu}$) contains no explicit direct 
dependence on the Riemann tensor, nonetheless there is actually an indirect 
dependence on the curvature in this particular case, since the second order 
differential equation obtained from the covariant Dirac equation is found to 
take the form $\nabla_{\mu}\nabla^{\mu}\psi(x)+
(1/4)R^{\alpha}_{\phantom {\alpha} \alpha}\psi(x)=0$, with the 
standard covariant coupling of a fermion to gravity thus always containing a 
non-inertial piece.\footnote{In passing we note when $\xi=1$ both the 
scalar and fermion wave equation operators can be written in the 
generic form $D_{\mu}D^{\mu}+(d/6)R^{\alpha}_{\phantom {\alpha} 
\alpha}$, where $d$ is the dimension of the field.} Likewise, even though the
curved space Maxwell equations, viz. $F^{\mu \nu}_{\phantom{\mu \nu};\nu}=0$, 
$F_{\mu \nu ; \lambda}+ F_{\lambda \mu ; \nu}+ F_{\nu \lambda ; \mu }=0$ 
also possess no direct coupling to curvature, in this case also there is a 
hidden dependence on curvature, with manipulation of the Maxwell equations 
leading to the second order equations 
$g^{\alpha \beta}F^{\mu\nu}_{\phantom {\mu\nu};\alpha; \beta}
+F^{\mu\alpha}R^{\nu}_{\phantom{\nu}\alpha}
-F^{\nu\alpha}R^{\mu}_{\phantom{\nu}\alpha}=0$
(or equivalently $g^{\alpha \beta} A_{\mu;\alpha; \beta}-
A^{\alpha}_{\phantom{\alpha};\alpha ;\mu}+
A^{\alpha}R_{\mu \alpha}=0$ where $A^{\mu}$ is the vector 
potential). Both the Dirac and Maxwell equations thus lead us\footnote{Because
fields are defined at all spacetime points and not just along the 
trajectories associated with point particles, the associated second order field 
equations are sensitive to the values of the Christoffel symbols at differing 
points. And even though it is possible to remove the Christoffel symbols at any
given point, it is nonetheless impossible to remove them from an entire region,
with the curvature dependent terms we find in the above second order wave
equations being the field theoretic generalization of the (covariantly 
describable) geodesic deviation found for pairs of nearby freely falling 
particles.} to an implicit expressly non-inertial coupling to the 
Riemann curvature tensor.\footnote{Since the only curvature dependence possible 
is in the form of couplings to the Ricci tensor, we note that any possible such 
terms are anyway immaterial to standard Ricci flat tests of the (weak) 
equivalence principle.} Thus rather than being quite unlikely, we see that an 
explicit coupling to curvature actually turns out to be the general rule in the 
field theoretic case, so that strictly geodesic behavior is not to ever be 
expected in general.\footnote{Thus we see that even if we do choose to generate 
general relativistic curved space field equations simply by covariantizing 
their special relativistic flat space forms, we still are led to non-inertial 
effects, and we are simply not free to postulate (viz. the weak equivalence 
principle) that those terms should be absent.} However, it turns out that in 
practice we can come very close to geodesic, and in order to see just how 
close, it is sufficient to explore the generic scalar field wave equation of 
Eq. (\ref{17}), a subject to which we now turn.

As regards Eq. (\ref{17}) we note immediately that if the scalar field has a 
non-zero mass, the ratio of the curvature term to the mass term in Eq. 
(\ref{17}) is given as $(L_C/L)^2$ where $L_C=\hbar/mc$ is the Compton 
wavelength of the scalar particle and $L$ is the scale on which 
the Ricci scalar varies ($R^{\alpha}_{\phantom {\alpha}\alpha}\sim 1/L^2$).
Thus if $L$ is macroscopic the effect of the Ricci scalar term will be 
completely irrelevant, and thus to explore any possible consequences due to 
the Ricci scalar term, we need only consider the massless limit of Eq. 
(\ref{17}). For it, we can again set $S(x)=$ exp$(iP(x)/\hbar)$, to obtain
\begin{equation}
P^{\mu}P_{\mu}-(\xi \hbar^2/6)R^{\alpha}_{\phantom {\alpha} \alpha}
=i\hbar P^{\mu}_{\phantom{\mu}; \mu}=0
\label{18}
\end{equation}
in the eikonal approximation. Equation (\ref{18}) admits of two types of 
solution depending on the strength of the Ricci scalar term. When the Ricci 
scalar term is weak we may set $P^{\mu}=\hbar dx^{\mu}/dq$, with $dx^{\mu}/dq$ 
then being found to be close to but not quite on the light cone (with the 
four-acceleration, unusually, not then being orthogonal to the four-velocity), 
with trajectories being found to obey
\begin{equation}
{d^2x^{\lambda} \over dq^2}
+\Gamma^{\lambda}_{\mu \nu}
{dx^{\mu} \over dq}{dx^{\nu } \over dq}={\xi \over 12}g^{\lambda \sigma}
(R^{\alpha}_{\phantom {\alpha} \alpha})_{; \sigma}.
\label{19}
\end{equation}
Since the right hand side of Eq. (\ref{19}) behaves as $1/L^3$, it will 
be negligible in the short wavelength limit unless $L$ is microscopic. Thus
in the short wavelength eikonal approximation we can nicely neglect the 
effects of any macroscopic Ricci scalar term. The second type of solution to 
Eq. (\ref{18}) is obtained when the Ricci scalar term is strong, a limit in 
which we may set    
\begin{equation}
P^{\mu}=\hbar (-\xi R^{\alpha}_{\phantom {\alpha} \alpha}/6)^{1/2} 
dx^{\mu}/d\tau
\label{20}
\end{equation}
a limit where the (thus microscopic) Ricci scalar now sets the scale for the 
quantum-mechanical wavelength. Given Eq. (\ref{20}), covariant differentiation 
of Eq. (\ref{18}) then directly yields
\begin{equation}
 R^{\beta}_{\phantom {\beta} \beta} 
\left( {d^2x^{\lambda} \over d\tau^2}
+\Gamma^{\lambda}_{\mu \nu}
{dx^{\mu} \over d\tau}{dx^{\nu } \over d\tau} \right)
=-\frac{1}{2} R^{\beta}_{\phantom {\beta} \beta ;\alpha}
\left( g^{\lambda \alpha}+                                                      
{dx^{\lambda} \over d\tau}                                                      
{dx^{\alpha} \over d\tau}\right).
\label{21}
\end{equation}
As we see, Eq. (\ref{21}) bears some resemblance to Eq. (\ref{10}), and indeed 
Eq. (\ref{21}) can also be obtained via variation of the point particle
action $I=-\kappa \int (R^{\alpha}_{\phantom {\alpha} \alpha})^{1/2} d\tau$,
where the coefficient $\kappa$ is actually dimensionless (in units of Planck's 
constant).\footnote{A point particle action in which the particle couples to 
the square root of the Ricci scalar is the only one available whose overall 
coefficient is dimensionless.} While we see that the motion associated with Eq. 
(\ref{21}) is not geodesic, such non-geodesic behavior would only be 
detectable microscopically with $L$ having to be of order the wavelength of 
the quantum-mechanical system.\footnote{In passing we note that in the 
microscopic case the Ricci scalar term acts somewhat like a mass term, and 
explicitly as one in fact in a background de Sitter geometry where 
$R^{\alpha}_{\phantom {\alpha} \alpha}$ takes the constant value $-12k$ and 
generates a mass $m=\hbar (2\xi k)^{1/2}/c$ to yield either a massive particle 
or a tachyon dependent on the sign of $\xi k$.} Thus if a large $L$, 
macroscopic, Ricci 
scalar term is to contribute in Eq. (\ref{17}) at all, it will only be able 
to do so in the long wavelength limit, a limit in which there would anyway be 
no eikonal approximation to begin with and in which we would (just as in the 
non-ray, $iT^{\mu}_{\phantom{;\mu};\mu}\neq 0$, regime of the standard 
classical Klein-Gordon equation) anyway have to use the full curved space 
Klein-Gordon or Schrodinger 
equations, with there then being no geodesic limit at all.\footnote{Thus in the 
presence of Eq. (\ref{17}) we either produce Eq. (\ref{4}) with no Eq. 
(\ref{10}) type terms, or we have to stay quantum-mechanical and never 
eikonalize at all.} Thus while the curvature term in Eq. (\ref{17}) might 
possibly have to be included in some quantum-mechanical situations, in those 
(short wavelength) cases where it is possible to make a ray approximation in
the first place, those rays will always be insensitive to the Ricci tensor 
dependent term, and will thus always lead to the geodesic equation of Eq. 
(\ref{4}) to very high accuracy.   

Thus, to conclude, we see that in this paper we have shown that the equivalence 
principle is indeed found to hold in quantum mechanics, that its  
primary characterization is as being purely geometric (our third version of the 
equivalence principle as given above), and that, moreover, it would appear that 
it is only through quantum mechanics that the classical-mechanical equivalence 
principle is even to be expected to hold at all. The author would like to thank 
W. Moreau, J. M. Bardeen and D. I. Santiago for helpful comments. This work has 
been supported in part by the 
Department of Energy under grant No. DE-FG02-92ER40716.00.

\vfill\eject

\end{document}